\documentclass[conference]{IEEEtran}
\IEEEoverridecommandlockouts

\usepackage[hidelinks]{hyperref}
\usepackage{amsmath,amssymb,amsfonts}
\usepackage{algorithmic}
\usepackage{graphicx}
\usepackage{textcomp}
\usepackage{xcolor}

\usepackage{siunitx} 
\usepackage[T1]{fontenc}
\usepackage{pmboxdraw}
\usepackage{caption}
\usepackage{subcaption}
\usepackage[table]{xcolor}
\usepackage{xspace}
\usepackage{csquotes}
\usepackage{fancyvrb}
\usepackage{makecell}
\usepackage{multirow}
\usepackage{nicematrix}
\usepackage{amsmath}
\usepackage[nameinlink]{cleveref}
\usepackage{booktabs}
\usepackage{wasysym} 
\usepackage[symbol]{footmisc}

\newcommand{\anonym}[2]{\ifx\myanonymous\undefined#1\else#2\fi}
\newcommand{\lichtenberg}{\anonym{Lichtenberg}{<AnonymousCluster>}}

\usepackage[backend=biber,style=ieee,maxcitenames=2,mincitenames=1,maxbibnames=99]{biblatex}
\renewcommand*{\bibfont}{\normalfont\small}

\crefname{equation}{Equation}{Equations}
\crefname{section}{Section}{Sections}
\crefname{figure}{Figure}{Figures}
\crefname{table}{Table}{Tables}

\def\BibTeX{{\rm B\kern-.05em{\sc i\kern-.025em b}\kern-.08em
    T\kern-.1667em\lower.7ex\hbox{E}\kern-.125emX}}

\DeclareMathOperator*{\argmin}{arg\,min}
\begin{document}

\title{Navigating the Energy Doldrums: Can We Exploit Energy-Price Volatility To Lower the Cost of Computing?
}

\author{\IEEEauthorblockN{Peter Arzt}
\IEEEauthorblockA{\textit{Department of Computer Science} \\
\textit{Technical University of Darmstadt}\\
Darmstadt, Germany \\
0000-0001-6937-1158}
\and
\IEEEauthorblockN{Felix Wolf}
\IEEEauthorblockA{\textit{Department of Computer Science} \\
\textit{Technical University of Darmstadt}\\
Darmstadt, Germany \\
0000-0001-6595-3599}
}

\maketitle

\begin{abstract}
Energy costs are a major factor in the total cost of ownership (TCO) for high-performance computing (HPC) systems.
The rise of intermittent green energy sources and reduced reliance on fossil fuels have introduced volatility into electricity markets, complicating energy budgeting. This paper explores variable capacity as a strategy for managing HPC energy costs -- dynamically adjusting compute resources in response to fluctuating electricity prices.
While this approach can lower energy expenses, it risks underutilizing costly hardware.
To evaluate this trade-off, we present a simple model that helps operators estimate the TCO impact of variable capacity strategies using key system parameters.
We apply this model to real data from a university HPC cluster and assess how different scenarios could affect the cost-effectiveness of this approach in the future.
\end{abstract}


\section{Introduction}
\label{sec:intro}
\begin{figure}
    \centering
    \includegraphics[width=\linewidth]{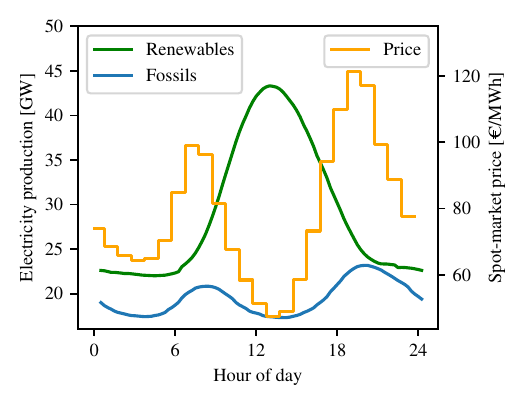}
    \caption{Electricity production and spot-market price in Germany over an average day in 2024. Data source: SMARD~\cite{SMARD}}
    \label{fig:diurnal-cycle}
\end{figure}

Energy costs represent a substantial component of the total cost of ownership (TCO) for high-performance computing (HPC) infrastructure, and have therefore consistently been a critical consideration in the design of large-scale computing facilities~\cite{sartor_money_2010, silva_review_2024, suarez_energy_2025}.
In recent years, however, the diminishing availability of cheap fossil energy -- driven in part by political shifts across Europe and broader efforts to reduce carbon emissions -- has introduced new complexities into the forecasting and management of energy budgets~\cite{cincinelli_role_2025}.
While green energy sources, such as solar and wind power, can often provide cheaper energy than conventional fossil-based or nuclear power plants~\cite{martin_de_lagarde_how_2018, paraschiv_impact_2014}, their inherent intermittency can pose challenges.
Output from these sources fluctuates with diurnal and seasonal cycles as well as weather conditions, making supply increasingly variable and difficult to predict.
This dynamic interplay between supply and demand, combined with limited energy storage capacity, has led to greater volatility in electricity prices.
For instance, prices can spike during periods of low solar and wind output (so-called \enquote{energy doldrums}), yet may also briefly fall below zero when generation vastly exceeds demand.
\cref{fig:diurnal-cycle} illustrates the average diurnal fluctuations in energy generation and their effect on the spot-market price of electricity.

One potential strategy for operators of high-performance computing (HPC) infrastructure to mitigate the impact of rising and volatile energy prices is to implement \textit{variable capacity}, where the amount of available compute power is adjusted according to current electricity prices~\cite{zhang_scheduling_2021}.
By reducing computing resources during periods of high energy costs, operators can lower energy consumption and thereby decrease overall expenses.
However, such shutdowns of compute infrastructure would leave hardware temporarily offline, reducing their utility in relation to the initial investment.
This presents a trade-off between improved energy efficiency and reduced hardware utilization.
To assess this trade-off, we propose a simple model to reason about the effect of volatile energy prices on the total cost of ownership of HPC systems and provide an upper bound of the economic viability of temporary shutdowns.
It is designed to act as a tool, aiding HPC operators in estimating whether temporary shutdowns could prove beneficial in their case, based on a small set of system parameters.
We demonstrate the application of this model using data from our university’s HPC cluster and explore how different future scenarios might influence the cost-effectiveness of variable capacity strategies.

The remainder of this paper is structured as follows: In \cref{sec:related-work}, we provide an overview of the existing literature on this topic.
Our model, designed to estimate the viability of temporary shutdowns to reduce energy costs, is introduced and visualized in \cref{sec:model}.
Then, we apply the model to a few scenarios in \cref{sec:scenarios}, before discussing its usefulness and limitations in \cref{sec:discussion}.
Finally, we conclude the paper in \cref{sec:conclusion-future-work}.

\section{Related Work}
\label{sec:related-work}
Dynamically adjusting the capacity of compute centers has been extensively studied in the literature.
Early work by \citeauthor{chase_managing_2001} and \citeauthor{pinheiro_load_2001} explored energy conservation through selectively powering servers on or off based on system load~\cite{chase_managing_2001, pinheiro_load_2001}.
In high-performance computing environments, however, variable capacity introduces additional constraints and complexity for resource scheduling.
Chien et al. formalized this scheduling problem~\cite{zhang_scheduling_2021} and examined strategies for coordination between datacenters and energy grid operators to reduce carbon emissions~\cite{lin_adapting_2023}.
More recently, researchers have proposed shutdown policies driven by reinforcement learning~\cite{casagrande_dont_2022}.

Variable capacity can only reduce greenhouse gas emissions and operating costs if shutdown policies account for a range of operational constraints.
These include the availability of renewable energy, limits on electricity consumption imposed by the energy grid, and, in particular, the time and energy overheads associated with shutting down and restarting nodes~\cite{benoit_reducing_2018}.
Even when these shutdown costs are considered, additional factors (CPU sleep states, potential impact on hardware lifetimes) have to be considered and addressed before variable capacity can be effectively implemented~\cite{rais_quantifying_2018}.

While existing literature often examines whether variable capacity can reduce carbon emissions or operating costs, it rarely addresses whether potential savings are economically viable in relation to the initial hardware investment.
In the following chapter, we present a model to address this gap.
Also investigating the balance between operating costs and hardware investment, although outside the context of high-performance computing, \citeauthor{bodner2025case} studied when the improved energy efficiency of modern hardware justifies replacing older systems~\cite{bodner2025case}.
The \enquote{Zero-Carbon Cloud} is a proposed concept to deploy servers in shipping containers close to renewable energy generation sites. Such a system could then provide cheap computing resources by acting as a dispatchable demand for excess renewable power that cannot be absorbed by the grid~\cite{chien_zero-carbon_2015}.
Our approach relies on a simplified price model aimed at quantifying price variability, whereas the field of \textit{electricity price forecasting} (EPF) focuses on modeling and predicting electricity markets using a wide range of mathematical techniques~\cite{lago_forecasting_2021}.

\section{Model}
\label{sec:model}
\newcommand{\plow}{p_\mathrm{low}}
\newcommand{\phigh}{p_\mathrm{high}}
\newcommand{\pavg}{p_\mathrm{avg}}
\newcommand{\pthresh}{p_\mathrm{thresh}}

We propose a simple model, designed to help operators of compute infrastructure to make estimations about the influence of fluctuating energy costs and reason about whether temporary shutdowns could be beneficial.
Working on an arbitrary, but fixed, time period $T$ about which we want to reason, we denote all costs caused by a given compute system in $T$ as its \textbf{total cost of ownership} $TCO$.
Furthermore, we assume that all expenses can be classified into two types of expenses: $TCO = F + E$.
\begin{enumerate}
    \item
    \textit{Fixed costs} $F$ are static costs that can not be lowered by shutdowns.
    Most notably, these cover write-offs for hardware procurement and other infrastructure costs (e.g., real estate, cooling hardware), but can also include expenditures for staff, administration and maintenance.

    \item
    \textit{Energy costs} $E$ are dynamic costs that can be reduced by (temporary) shutdowns of the compute infrastructure.
    Primarily, these consist of the electricity costs for operating the system, but could also include other costs that can be reduced by variable capacity (e.g., consumable supplies).
\end{enumerate}
The classification into fixed costs and energy costs roughly aligns with the concepts of \textit{capital expenditure} (CapEx) and \textit{operational expenditure} (OpEx), respectively.

\noindent
\paragraph{Price model}
We use the following approach to model the volatility of energy prices:
Given the energy prices during $T$, sampled in some regular interval, as $p_{1...n}$ and their overall average as $\pavg$, we define two distinct price regions:
All prices exceeding a threshold are called high, all others are low.
While the low-price region encompasses low prices observed during normal operations, the high-price region is meant to represent conditions under which temporary shutdowns may become economically justified.
We denote by $x\in(0,1)$ the fraction of the time period during which prices fall within the high-price region.
Hence, we can compute the threshold price $p_{thresh}$ that separates the two regions using $p_{1...n}$ and $x$:
\begin{align}
    \label{eq:threshold}
    &p_{thresh} = Q_{(1-x)}(p_{1...n}), \\
    &\;\;\text{where $Q_p(p_{1...n})$ is the p-percentile of samples $p_{1...n}$}. \nonumber
\end{align}

\noindent
Furthermore, we define $\plow$ and $\phigh$ as the average prices in their respective price regions, so that $\pavg$ can be expressed as their weighted mean:
\begin{equation}
    \pavg = x \cdot \phigh + (1-x) \cdot \plow
    \label{eq:phigh-plow}
\end{equation}

\begin{figure*}[tb]
    \centering
    \includegraphics[width=\linewidth]{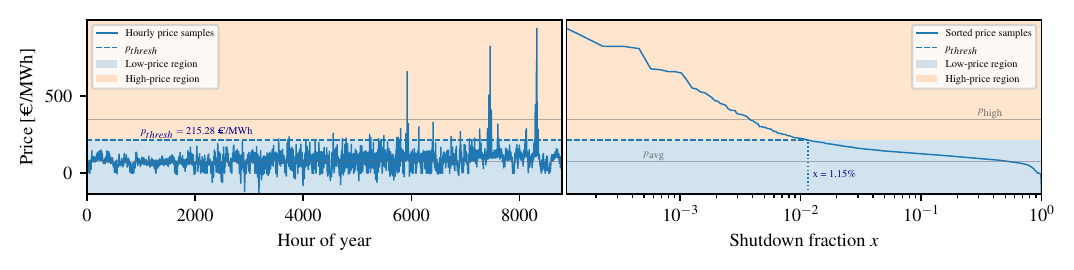}
    \caption{Visualization of our model for energy prices, demonstrated using Germany's historic day-ahead prices from 2024 (data source: SMARD~\cite{SMARD}, resolution: 1 hour).
    The left diagram plots the prices chronologically, while the right one depicts the price samples in descending order over a logarithmic x-axis, akin to a \textit{survival function} from statistics.
    For a given shutdown fraction $x$ (e.g., $x = 1.15\%$), the price threshold is determined as defined in \cref{eq:threshold}, which is then used to categorize the price samples into regions of low and high prices, illustrated by the blue and orange areas.}
    \label{fig:energy-price-model-viz}
\end{figure*}

\noindent
Finally, $k$ is the relative price increase during high-price phases compared to average prices:
\begin{align}
    k &= \frac{\phigh}{\pavg}, \text{with $k > 1$}
    \label{eq:k}
\end{align}
Combining this with \cref{eq:phigh-plow}, we can express $\plow$ and $\phigh$ based on $\pavg$, $k$ and $x$:
\begin{align}
    \label{eq:phigh-as-function}
    \phigh &= \pavg \cdot k \\
    \plow &= \pavg \cdot \frac{kx - 1}{x - 1}
    \label{eq:plow-as-function}
\end{align}
\cref{fig:energy-price-model-viz} visualizes the model we use to model energy prices and \cref{tab:symbol-overview} provides an overview of the different model parameters and symbols used in this chapter.

\renewcommand{\theadfont}{\normalsize\bfseries}
\begin{table}[tb]
    \centering
\begin{NiceTabular}{l|l|l } 
\CodeBefore
  \rowcolor{gray!50}{1}
  \rowcolors{2}{gray!20}{}
\Body
  \toprule
  \thead{Symbol} & \thead{Definition} & \thead{Unit} \\
  \midrule
    TCO & Total cost of ownership & \euro \\
    $T$ & Time period & hour\\
    $F$ & Fixed costs & \euro \\
    $E$ & Energy costs & \euro \\
    $p_{1,...n}$ & Energy prices, sampled over T & \euro/MWh \\
    $p_{thresh}$ & Threshold price & \euro/MWh \\
    $\pavg$ & Average energy prices over $T$ & \euro/MWh\\
    $\phigh$ & \Block{}{Average energy prices\\while in high-price region} & \euro/MWh \\
    $\plow$ & \Block{}{Average energy prices\\while in low-price region} & \euro/MWh\\
    $k$ & $\frac{\phigh}{\pavg}$ & -\\
    $x$ & \Block{}{Shutdown fraction: fraction of time\\with high energy prices / \\ with system shutdown, $x \in (0; 1)$} & -\\
    CPC & Cost per compute & \euro \\
    $C$ & Power consumption under full operation & MW \\
    $\Psi$ & Cost distribution coefficient $\frac{F}{T \cdot C \cdot \pavg}$ & -\\
  \bottomrule
\end{NiceTabular}
\caption{Overview of model parameters and symbols. Units are listed to exemplify the parameter's dimension and can be replaced by other units with equivalent dimensionality.}
\label{tab:symbol-overview}
\end{table}

\paragraph{Shutdown policies}
We consider two fundamental shutdown policies and describe their overall energy costs based on the parameters $\pavg$, $k$, $x$, $T$, and the system power consumption under full operation~$C$.
\enquote{Always on} (AO) describes a no-shutdown policy where no temporary shutdowns are performed, independent of the energy price.
\enquote{With shutdowns} (WS), on the other hand, means that the whole infrastructure is shut down during phases with high energy prices, leaving only $\plow$ relevant for the overall energy costs:
\begin{align}
    E_\text{AO} &= T \cdot C \cdot \pavg\\
    E_\text{WS} &= T \cdot C\cdot (1 - x) \cdot \plow \\
                    &= T \cdot C \cdot (1 - x) \cdot \pavg \cdot \frac{k x - 1}{x - 1} \\
                    &= T \cdot C \cdot \pavg \cdot (1 - k x)
\end{align}

\noindent
To compare the scenarios of shutdowns during high energy price phases against no shutdowns, we consider the quotient of the TCO and the time that the system is operational and denote it as the \textit{cost~per~compute}~(CPC).
The ratios for the two policies are given as follows:
\begin{align}
    \text{CPC}_\text{AO} &= \frac{F + E_\text{AO}}{T} \\
                                    &= \frac{F + T \cdot C \cdot \pavg}{T} \\
    \text{CPC}_\text{WS} &= \frac{F + E_\text{WS}}{(1 - x) \cdot T} \\
                                    &= \frac{F + T \cdot C \cdot \pavg \cdot (1 - k x)}{(1 - x) \cdot T}
\end{align}

\noindent
Working with all the above assumptions, we can thus express the question, whether shutdowns are beneficial for cost efficiency, given $F$, $\pavg$, $k$ and $x$, as a simple inequality:
\begin{align}
    \text{CPC}_\text{WS} &< \text{CPC}_\text{AO} \\
    \Longleftrightarrow\;\;\;\frac{F + T \cdot C \cdot \pavg \cdot (1 - k x)}{(1 - x) \cdot T} &< \frac{F + T \cdot C \cdot \pavg}{T} \\
    \Longleftrightarrow\;\;\;k &> \frac{F}{T \cdot C \cdot \pavg} + 1 \\
    \Longleftrightarrow\;\;\;k &> \frac{F}{E_\text{AO}} + 1    
\end{align}

\noindent
Interestingly, this inequality does not depend on $x$, but only relies on $k$ and $\pavg$ to characterize energy prices.
The ratio $\frac{F}{E_\text{AO}}$ reflects the cost distribution of the compute system, which we consider to be an inherent property of the system and denote it by~$\Psi$:
\begin{align}
    \Psi &= \frac{F}{E_\text{AO}} = \frac{F}{T \cdot C \cdot \pavg}
\end{align}

\noindent
This brings us to the following concise inequality, which expresses the model's prediction whether temporary shutdowns are economically viable:
\begin{equation}
    k > \Psi + 1
    \label{eq:model}
\end{equation}

\section{Case Studies}
\label{sec:scenarios}
\newcommand{\coo}{\ensuremath{\mathrm{CO_2}}}

To demonstrate how our model can be used to reason about the efficacy of temporary shutdowns, we apply it to a selection of cases.
We start by considering our university's high-performance computer \enquote{\lichtenberg}, before moving on to more hypothetical scenarios.

\subsection{\lichtenberg{}}
\label{subsec:lichtenberg}
To model \lichtenberg{}'s cost distribution, we first estimate the parameter~$\Psi$.
Since the cluster comprises multiple partitions built with different hardware generations, whose operational lifetimes typically overlap by a few years, we assume a yearly average for hardware procurement costs.
In addition to hardware costs, we account for expenses related to the building infrastructure and cooling systems, which are long-term investments spanning multiple cluster generations.
The sum of these components constitutes our estimate of the fixed costs.
By dividing the fixed costs by the average annual electricity expenditure, we obtain an estimated cost distribution coefficient of approximately $\Psi_{LB} \approx 2$, independent of $T$ and $C$.

We need to model the fluctuations in electricity prices using the price model described in \cref{sec:model}.
Although in reality, \lichtenberg{} currently draws power on a fixed-price contract, we assume here that its energy costs would be determined by Germany's spot-market electricity prices.
Thus, we use the historical day-ahead prices from 2024 (data source: SMARD~\cite{SMARD}) with varying sampling intervals and different time scales to model price fluctuations on different time scales.
For all possible values of $x \in (0; 1)$, we compute the factor $k$ as described in \cref{eq:threshold,eq:phigh-plow,eq:k,eq:phigh-as-function,eq:plow-as-function}, giving us a set of $(k, x)$-pairs that describe the \textit{price variability} $PV$ of $p_{1...n}$:
\begin{equation}
    PV = \left\{ (k, x) : k = \frac{\phigh}{\pavg}\right\}
    \label{fig:pv}
\end{equation}

\begin{figure}[tb]
    \centering
    \includegraphics[width=0.99\linewidth]{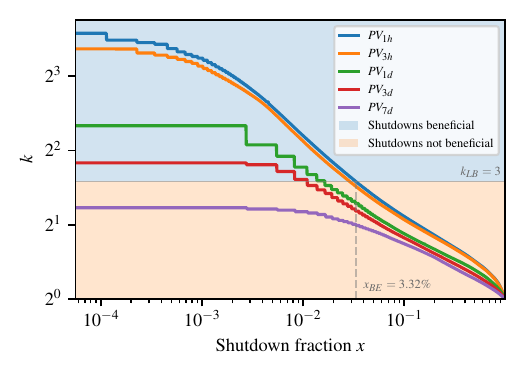}
    \caption{
    Price variability $PV$ of Germany's spot-market electricity prices from 2024~\cite{SMARD}, for different sampling intervals, as defined in \cref{fig:pv}.
    The blue area depicts the range of $k$ for which our model assesses temporary shutdowns to be viable, based on our model in \cref{eq:model} and our estimation for \lichtenberg{}'s $\Psi_{LB} = 2$.
    The point where a k-x-line leaves the blue area represents the point after which shutdowns are no longer beneficial (break-even point $x_{BE}$).
    For weekly samples, the model predicts that shutdowns are counterproductive in every case.
    For $PV_{1h}$, the model predicts that shutdowns are beneficial when $x < 3.32\%$.
    Note that both axes are logarithmic.
    }
    \label{fig:kx}
\end{figure}

\noindent
The resulting set can then be plotted as a \textit{k-x line} for electricity prices sampled using different intervals, as depicted in \cref{fig:kx}.
For weekly price samples, $k$ does not exceed $k_{LB} = 3$ for all $x$.
Thus, the model predicts that temporary shutdowns lasting about 7 days are never beneficial, regardless of the chosen threshold price.
However, the picture is different on shorter time scales.
Using one hour as the sampling interval, the k-x-line leaves the beneficial zone at $x_{BE} = 3.32\%$, indicating that shutting down the system up to $3.32\%$ of the time would be beneficial.

In the next step, we can use our model to estimate the ideal value for $x$, i.e., the optimal fraction of time that the system should be turned off to minimize CPC:

\begin{align}
    x_{opt} &= \argmin_{(k, x) \in PV_{1h}} CPC_{WS} \\
    &= \argmin_{(k, x) \in PV_{1h}} \frac{F + T \cdot P \cdot \pavg \cdot (1 - kx)}{(1-x) \cdot T \cdot P} \\
    &= \argmin_{(k, x) \in PV_{1h}} \frac{1 - kx + \Psi_{LB}}{1 - x} \\
    &= 0.8189 \% \\
    k_{opt} &= 4.9726
\end{align}

Hence, the model predicts that it is optimal in terms of cost efficiency to shut down the system for approximately $0.8\%$ of the time.
Finally, we can estimate the relative CPC reduction, compared to performing no shutdowns at all:

\begin{align}
    &  1 - \frac{\text{CPC}_\text{WS}}{\text{CPC}_\text{AO}} \\
    &= 1- \frac{\left( \frac{F + T \cdot P \cdot \pavg \cdot (1-k_{opt}x_{opt})}{(1-x_{opt}) \cdot T \cdot P} \right)}{\left( \frac{F + T \cdot P \cdot \pavg}{T \cdot P}\right)} \\
    &= 1 - \frac{\Psi + 1 - k_{opt} \cdot x_{opt}}{(\Psi + 1)(1 - x)} \\
    &= 0.5429 \%
\end{align}

\noindent
Overall, the model predicts that, with an optimal shutdown policy that disables the system for about $0.8\%$ of the overall time, and under all model assumption that we will discuss in \cref{sec:discussion}, it would be possible to improve CPC by around $0.54\%$ by performing shutdowns of at least one hour during phases of high energy prices.
The corresponding threshold price for shutdowns is $237.84$ \euro/MWh.

\subsection{Increased price variability (South Australia)}
\label{subsec:south-australia}
\begin{figure}
    \centering
    \includegraphics[width=\linewidth]{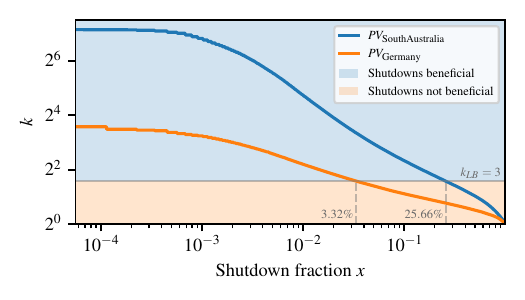}
    \caption{Comparison of the respective price variabilities $PV$ of hourly intra-day spot-market prices (Australia: \textit{dispatch price}) of Germany~\cite{SMARD} and South Australia~\cite{AEMO}. Similar to \cref{fig:kx}, the orange and blue areas represent the models viability prediction for $\Psi = 2$ and the hatched lines mark the respective break-even points after which shutdowns are no longer beneficial. Both axes are logarithmic.}
    \label{fig:germany-vs-south-australia}
\end{figure}

In the next step, we can use the model to investigate more hypothetical scenarios.
In general, there are two main factors that determine the model's assessment of economic viability: The degree of price variability (modeled using $k$ and $x$) and the system cost distribution $\Psi$.
To examine the effects of increased price variability, we consider a system with a cost distribution comparable to that of \lichtenberg{} ($\Psi = 2$), operating in the South Australian electricity market, where price fluctuations are among the highest globally~\cite{CORNELL20241421}.
\cref{fig:germany-vs-south-australia} illustrates the price variability in 2024 compared to Germany's electricity market.
Analogously to the previous section, we use the 2024 price samples (data source: AEMO~\cite{AEMO}) as input for our model and compute the break-even point, after which temporary shutdowns are no longer beneficial:
The results show that, compared to the situation in the German electricity market where the break-even point is at $x_{BE} = 3.32\%$, the model predicts that shutdowns up to $25.66\%$ of the time are beneficial.
A similar result can be observed for the most efficient shutdown configuration: The ideal shutdown fraction $x$ more than quadruples to $x_{opt} = 3.66\%$ which would result in a theoretical CPC reduction of $8.31\%$ over a no-shutdown policy.

\subsection{Shifted distribution of costs}
Besides increased price variability, the second factor that can affect the economic viability of temporary shutdowns is a shift in the cost distribution between fixed costs and energy costs, which we denote by $\Psi$.
Such a shift may occur in the future due to one or a combination of the following reasons:
Should the average electricity price rise in the future while hardware costs stay roughly constant, $\Psi$ would decrease, thus making temporary shutdowns more likely to be beneficial (c.f.,~\cref{eq:model}).
Analogously, decreased hardware procurement expenditure (with constant electricity costs) would have the same effect.

\begin{figure}
    \centering
    \includegraphics[width=\linewidth]{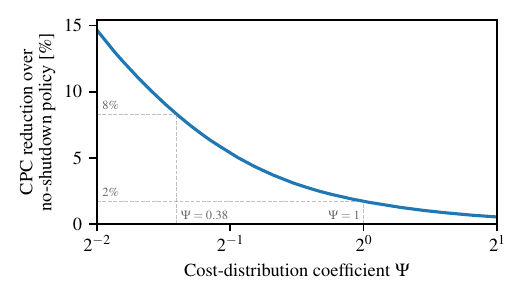}
    \caption{Maximum theoretical CPC reduction of temporary shutdowns over a no-shutdown policy for varying values for the cost-distribution coefficient $\Psi$, assuming the Germany's 2024 historic prices (resolution: 1 hour). The x-axis is logarithmic.}
    \label{fig:cheaper-hardware}
\end{figure}

To quantify the influence of $\Psi$ on the maximum theoretical benefit of shutdowns, we compute the CPC reduction of an optimal shutdown configuration over a no-shutdown policy for varying values of $\Psi$.
The results, depicted in \cref{fig:cheaper-hardware}, show that the fraction between fixed costs and energy costs would need to fall to $\Psi = 0.38$ to achieve a CPC reduction that is comparable to our scenario with South Australia's price variability.
Compared to \lichtenberg{}'s cost distribution ($\Psi_{LB} = 2$), hardware expenses would thus need to be reduced by approximately $\sim 81\%$ to reduce CPC by $\sim 8\%$ compared to a no-shutdown policy.

\subsection{Combined scenario}
\label{subsec:combined-scenario}

Combining both previously discussed aspects, we look at a scenario in which multiple model parameters are changed.
Specifically, starting from the scenario in \cref{subsec:lichtenberg}, we consider the following two developments which are motivated by real-world trends:
\begin{enumerate}
    \item
    To simulate higher taxes on carbon emissions and renewable energy prices that continue to fall, we use artificially generated price samples with increased variability.
    To produce this hypothetical price data, we use the same spot-market electricity prices as in \cref{subsec:lichtenberg} (resolution: 1 hour), but apply a scaling factor to all non-negative price samples that is determined by the fraction of fossil energy production at that time.
    In more formal terms, based on the original prices $p_{1,...,n}$ and the fossil and renewable energy production volumes\footnote{We assume here that the overall electricity generation can be categorized into fossil and renewable sources. Renewables cover wind  (onshore and offshore), solar and biomass energy, while fossils include coal and gas energy. Nuclear energy is not considered as Germany shut down its last two reactors in 2023.} $p^{\text{fossil}}_{1,...,n} / p^{\text{renewable}}_{1,...,n}$, we define our new prices $\widetilde{p_{1,...,n}}$ as:
    \begin{align}
        \widetilde{p_i} = \begin{cases}
            p_i, & \text{if $p_i \leq 0$} \\
            \frac{p_i \cdot (1-\beta_i)}{2} + p_i \cdot \beta_i \cdot 2, & \text{else,} \\
        \end{cases} \\
        \text{where}\; \beta_i = \frac{p^{\text{fossil}}_i}{p^{\text{fossil}}_i + p^{\text{renewable}}_i} \text{.}\nonumber
    \end{align}

    \item We speculate that the fixed costs of a future system are reduced by $20\%$, e.g., due to falling hardware prices.
\end{enumerate}

\begin{figure}[tb]
    \centering
    \includegraphics[width=\linewidth]{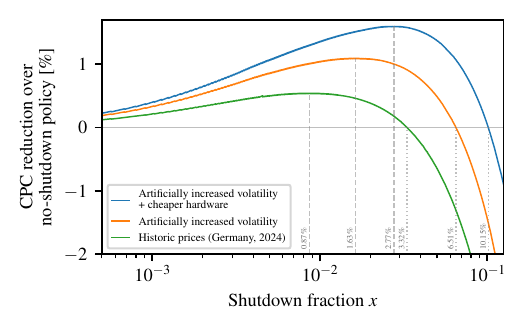}
    \caption{
    Illustration of the trade-off between energy cost saving and reduced hardware utilization for historic prices (Germany's spot-market, 2024, resolution: day) and two hypothetical scenarios: Increased price volatility and, additionally, comparatively lower hardware prices.
    The break-even point and optimal configurations are marked for each of the curves.
    Different shutdown fractions $x$ are represented on the logarithmic x-axis, while the y-axis plots the respective CPC reductions compared to a no-shutdown policy.
    }
    \label{fig:combined-scenario}
\end{figure}

While the first trend only affects the price model, the reduced fixed costs lead to an updated cost-distribution coefficient of $\Psi = 1.6$.
We then apply our model to assess the economic viability in this hypothetical setup.
\cref{fig:combined-scenario} illustrates the trade-off between energy cost savings and reduced hardware utilization that is being assessed by the model.
The results show that a combination of increased price variability and cheaper hardware procurement makes temporary shutdowns economically viable for larger time portions.
In the combined scenario, the model estimates that shutdowns are viable up to $10.15\%$ of the time, with the predicted optimum being at shutting down the system $2.77\%$ of the time.

\subsection{Regional comparison}
\label{subsec:regional-comparison}
\begin{figure}
    \centering
    \includegraphics[width=0.99\linewidth]{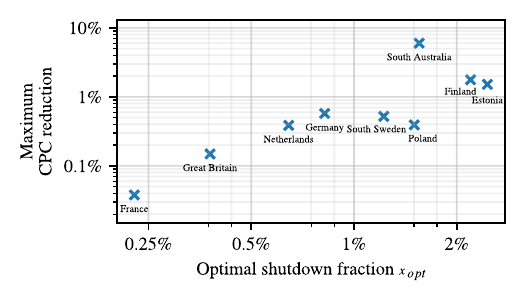}
    \caption{Illustration of the optimal shutdown configurations and respective predicted CPC reductions for the geographical regions considered in \cref{subsec:regional-comparison} and \cref{tab:region-comparison}. The x-axis is logarithmic.}
    \label{fig:country-comparison}
\end{figure}

To assess the economic viability of temporary shutdowns across different geographic regions, we apply our model to a selection of regions with available historic price data (data source: Electricity~Maps~\cite{electricitymaps}, resolution: 1 hour). 
For each region, we calculate the average price $\pavg$ \footnote{Non-euro prices were converted to Euros using conversion rates queries from the \textit{Frankfurter} API (\url{https://api.frankfurter.dev/}, last accessed on 2025-12-02)} and determined the corresponding cost-distribution coefficient ($\Psi$), assuming the system's fixed costs and electricity consumption are equivalent to those of \lichtenberg{}.
Thus, we effectively simulate hypothetical scenarios in which the cluster operates across various energy markets.
Our model yields estimates of the break-even and optimal shutdown fractions ($x$) for each region, which are summarized in \cref{tab:region-comparison}.
Furthermore, \cref{fig:country-comparison} illustrates the optimal shutdown configurations and theoretical CPC reductions relative to a no-shutdown policy.
The results reveal a substantial range of predicted maximum CPC reductions, with notable variations across regions. 
Specifically, our analysis suggests that shutdowns are not viable for Spain at the selected data resolution, whereas South Australia is predicted to achieve a maximum CPC reduction exceeding \SI{5}{\percent}\footnote{These results for South Australia differ from those presented in \cref{subsec:south-australia} due to differences in the value chosen for $\Psi$.}.

\renewcommand{\theadfont}{\normalsize\bfseries}
\begin{table}[tb]
    \centering
\begin{NiceTabular}{l|r|r|r|rr} 
\CodeBefore
  \rowcolor{gray!50}{1}
  \rowcolors{2}{gray!20}{}
\Body
\toprule
\thead{Region} & \thead{$\pavg$} & \thead{$\Psi$} & \thead{$x_{BE}$} & \thead{$x_{opt}$} & \thead{CPC red.} \\
\rowcolor{gray!50}
& [\euro/MWh] & & [\%] & [\%] & [\%] \\
\midrule
South Australia & \num{59.36} & \num{2.62} & \num{17.55} & \num{1.55} & \num{5.99}\\
Finland & \num{46.36} & \num{3.36} & \num{8.25} & \num{2.20} & \num{1.76}\\
Estonia & \num{87.69} & \num{1.77} & \num{9.24} & \num{2.46} & \num{1.52}\\
Germany & \num{77.84} & \num{2.00} & \num{3.34} & \num{0.82} & \num{0.57}\\
South Sweden & \num{50.05} & \num{3.11} & \num{3.75} & \num{1.22} & \num{0.52}\\
Poland & \num{96.26} & \num{1.62} & \num{4.04} & \num{1.50} & \num{0.39}\\
Netherlands & \num{77.60} & \num{2.01} & \num{2.54} & \num{0.64} & \num{0.39}\\
Great Britain & \num{85.92} & \num{1.81} & \num{1.12} & \num{0.38} & \num{0.15}\\
France & \num{58.19} & \num{2.67} & \num{0.53} & \num{0.23} & \num{0.04}\\
Spain & \num{63.09} & \num{2.47} & - & - & -\\

\bottomrule
\end{NiceTabular}
\caption{Results of the regional comparison in \cref{subsec:regional-comparison}. Regions are sorted by the predicted maximum efficiency increase (rightmost column) in descending order. The column titled \textit{CPC red.} lists the maximum CPC reduction, relative to a no-shutdown policy.}
\label{tab:region-comparison}
\end{table}

\section{Discussion}
\label{sec:discussion}
The model introduced in \cref{sec:model} and demonstrated in \cref{sec:scenarios} is intended to provide a rough estimate of whether temporary shutdowns can serve as an effective strategy for managing energy costs amid increasing energy price volatility.
Before evaluating its predictions in a real-world context (\cref{subsec:case-study-results}) and reflecting challenges that restrain the implementation of variable capacity (\cref{subsec:hurdles}), it is essential to consider the model’s underlying assumptions and limitations (\cref{subsec:model-limitations}).

\subsection{Model limitations}
\label{subsec:model-limitations}
The model abstracts from several real-world complexities.
A central assumption is that, when the compute cluster is not undergoing a temporary shutdown, its hardware is fully utilized.
In other words, we assume that every available compute hour is used productively.
This simplification ignores potential underutilization due to technical or administrative factors, such as scheduling gaps or decreased demand during periods like holiday breaks.
Although this does not accurately reflect the typical utilization patterns of most real-world clusters, it is a deliberate simplification intended to isolate the effects of energy price volatility.
In reality, clusters with significant underutilization may already be viable candidates for temporary shutdowns, independent of energy market dynamics.

\paragraph{Shutdown costs}
The model further assumes that temporary shutdowns can be executed instantaneously and without cost, a simplification that does not reflect the behavior of most real-world computing hardware.
In practice, cluster nodes typically require non-negligible time and energy to shut down and restart~\cite{benoit_reducing_2018}.
Such costs are excluded from the model because they are difficult to estimate, as they depend on specific hardware characteristics and operational policies.
However, this omission introduces a bias, favoring shutdowns by overstating their potential benefits.
As a result, the model's predictions regarding the economic viability of temporary shutdowns should be interpreted as an upper bound, likely overestimating their cost-effectiveness.

This bias reduces the reliability of the model’s predictions when it deems temporary shutdowns economically viable, but enhances confidence in its assessments when shutdowns are predicted to be unprofitable.
Applied to our case study on the \anonym{Lichtenberg cluster}{\lichtenberg} (\cref{subsec:lichtenberg}), this implies that although the model indicates a potential benefit from temporary shutdowns based on 2024 spot-market prices (with a resolution of 1 hour), the predicted reduction in cost-per-compute (CPC) is minimal ($0.54\%$).
Given this small margin, it is unlikely that shutdowns would have yielded a meaningful improvement in cost efficiency once real-world shutdown costs are taken into account.

\paragraph{Optimization criterion and constraints}
The model is centered on the concept of cost-per-compute as its primary optimization criterion.
While cost efficiency is a key concern for operators of compute infrastructure, it is typically only one of several relevant factors.
Depending on the specific objectives, the model can be adapted to optimize for alternative criteria.
For instance, in efforts to minimize a cluster’s carbon footprint, financial costs could be replaced with estimated carbon emissions, enabling analysis in terms of emissions-per-compute.
However, in practice, cluster operators must also account for a variety of additional objectives and constraints that are not easily captured within the model.
For example, high electricity prices might force operators with tight budgets to temporarily shut down hardware, even at the expense of reduced cost efficiency.
Conversely, operators may opt to avoid shutdowns -- despite potential cost savings -- in order to maintain service-level goals such as short job queue times.

\paragraph{Partial shutdowns}
The model considers only two operational states: full utilization during periods of low energy prices and complete shutdown during periods of high energy prices.
In practice, however, full shutdowns would likely only be a last resort for cluster operators, who may prefer more flexible strategies such as partial shutdowns.
Nonetheless, within the framework of our model, partial shutdowns cannot emerge as optimal.
Assuming a homogeneous system architecture, the trade-off between energy cost savings and reduced hardware utilization is identical for any subsystem as it is for the entire cluster. As a result, the model will always favor either full operation or complete shutdown.

This limitation does not imply that partial shutdowns are without merit in real-world settings.
Additional objectives and constraints that are beyond the scope of the model can make partial shutdowns viable.
For instance, during periods of very high energy prices when the model recommends a full shutdown, an operator might choose to maintain a subset of the cluster to ensure basic service availability.
Similarly, heterogeneous clusters composed of partitions with varying energy efficiencies could benefit from selectively shutting down the most energy-intensive components.
Our model could then be employed for cluster partitions to assess the viability of shutdowns individually.

\paragraph{Negative electricity prices}
As depicted in \cref{fig:energy-price-model-viz}, phases with negative electricity prices are not uncommon occurrences in some markets.
In the context of our model, this has no implications on the prediction validity as long as the average price $\pavg$ remains positive.
If $\pavg$ were negative, the model would no longer be applicable, since the definition of $k$ would become invalid, c.f.,~\cref{eq:k}.
This, however, does not diminish the model’s real-world relevance, as persistently negative electricity prices would imply a fundamentally inverted energy market, a scenario that appears highly unrealistic in the foreseeable future and would render considerations of the economic viability of variable capacity irrelevant.

\subsection{Case study results}
\label{subsec:case-study-results}
In \cref{sec:scenarios}, we apply our model to \lichtenberg{}, the HPC cluster at our university.
We utilize spot-market electricity prices from 2024 to model historical price fluctuations, alongside a hypothetical scenario featuring artificially increased price volatility.
As illustrated in \cref{fig:kx}, the model’s assessment is sensitive to the temporal resolution of the price data: lower-resolution price samples inherently smooth out short-term spikes, reducing observed variability.
Therefore, analysts using the model to evaluate the viability of temporary shutdowns should select price data with a resolution that corresponds to the shutdown timescales under consideration.
For instance, a cluster capable of disabling and restarting nodes within minutes may benefit from short shutdowns lasting minutes or hours, whereas less flexible systems require longer-term price fluctuations to realize gains from variable capacity.

Overall, the case studies indicate that, based on historical price fluctuations, temporary shutdowns of the \anonym{Lichtenberg cluster}{\lichtenberg} in 2024 could not have resulted in more than marginal efficiency improvements.
However, when considering price data from the South Australian electricity market, which serves as an example of extreme real-world price variability, it becomes evident that there are scenarios in which variable capacity can yield substantial improvements in cost efficiency.

In addition to price variability, the model’s assessment is sensitive to the relative weight of energy expenses compared to fixed costs, represented by the cost-distribution coefficient~$\Psi$.
Several factors could cause the cost coefficient to fall in the future, making shutdowns more likely to be viable:
For instance, a sustained rise in energy prices without a corresponding increase in hardware costs would increase the relative impact of operational expenses.
Additionally, longer hardware lifetimes (due to less frequent upgrades) would distribute hardware costs over a longer period, effectively lowering fixed costs.
Market changes, such as increased competition among GPU vendors, could also reduce hardware prices by lowering profit margins.

Another aspect that influences $\Psi$ and could therefore tip the scales for the economic viability of variable capacity are economies of scale.
Even though the specific cases are difficult to assess without detailed financial information, we can assume that operators of very large compute infrastructure, ranging up to Tier-0 supercomputers and cloud hyperscalers, can procure hardware at discounted prices and spend less money on staff and maintenance per compute hour.
The resulting lower cost-distribution coefficient $\Psi$ would make temporary shutdowns beneficial in more cases compared to smaller systems, suggesting that variable-capacity operation may become economically attractive first for large-scale operators rather than for small or medium-sized clusters.

As their effects can accumulate, as illustrated in \cref{fig:combined-scenario}, a combination of factors influencing both price variability and the relative weight of energy costs ($\Psi$) could strengthen the economic case for temporary shutdowns in the future.
However, shifts in these key parameters might also enable a different approach: the procurement of larger, intentionally overprovisioned systems.
Such systems would be designed with variable-capacity operation in mind, running at full capacity during periods of low energy prices and partially or fully scaling down when energy costs are high.
Higher upfront procurement costs could be amortized by the increased cost efficiency achieved through flexible operation, enabling operators to optimize energy consumption without compromising peak performance during favorable conditions.

\paragraph{Regional comparison}
The analysis in \cref{subsec:regional-comparison} shows that the economic viability of temporary shutdowns is sensitive to the geographic location.
While there is no economic case for variable capacity in countries with stable and comparatively low electricity prices (e.g., Spain and France), the predicted upper bounds for efficiency increases over a no-shutdown policy are higher in regions with higher price variability.

However, the analysis makes a range of simplifying assumptions, limiting its real-world validity:
The comparison is based on data from the spot market for electricity, i.e., on wholesale prices which exclude taxes, network fees, etc.
As these additional costs vary between countries, the model predictions would change when considering retail prices.
Furthermore, the assumptions that fixed costs are independent of the system's geographical location is unlikely to be accurate in the real-world.
Still, the presented analysis demonstrates that significant differences exist between geographic regions, making it indispensable to consider the dynamics of the relevant energy market when assessing variable capacity.

\subsection{Hurdles for variable capacity}
\label{subsec:hurdles}
Aside from the question of economic viability, there are also administrative and technical hurdles to the implementation of variable capacity.
In the case of the \anonym{Lichtenberg cluster}{\lichtenberg}, which currently operates using electricity purchased through a fixed-price contract, increasing cost efficiency using variable capacity would require switching to an electricity contract that reflects the fluctuations of the spot market for electricity.
Also, today's HPC software stacks (e.g., communication libraries like MPI and schedulers like SLURM) are currently not designed with variable capacity in mind.
Efforts for more flexible software stacks are underway~\cite{tarraf_malleability_2024, iserte_efficient_2017}, but are still in the process of reaching maturity.

\section{Conclusion and Future Directions}
\label{sec:conclusion-future-work}
Managing energy costs and reducing carbon emissions have become increasingly important for operators of high-performance computing (HPC) infrastructure, particularly in the face of growing electricity price volatility.
Variable capacity, i.e., dynamically adjusting a system’s computational resources in response to factors such as electricity prices or demand, offers a promising means of adapting traditionally rigid HPC clusters to this changing environment.
However, shutdown policies aimed at reducing energy consumption during periods of high prices inevitably involve a trade-off between cost savings and reduced hardware utilization.
Additional operational constraints, extending beyond cost efficiency, often introduce further complexity.
In this work, we introduced a simple model that relates potential energy savings to the initial hardware investment, enabling the identification of a balance between energy conservation and underutilization.
Using key system parameters and hypothetical electricity price data, the model estimates an upper bound on the economic viability of temporary shutdowns during high-price periods.

We applied our model to our university’s HPC cluster, \lichtenberg{}, using historical price data from Germany’s electricity spot market.
For 2024, the model estimated that shutting down the cluster for approximately $0.8\%$ of the time could have increased energy efficiency by $0.54\%$.
When incorporating additional constraints, however, it appears unlikely that temporary shutdowns would have yielded net efficiency gains in 2024.
Nonetheless, hypothetical case studies illustrate that the theoretical efficiency gains vary between geographic regions and that greater electricity price variability or comparatively lower hardware costs could enhance the economic viability of variable capacity in the future.
Furthermore, our model implies that variable-capacity operation will become economically attractive first for large-scale operators such as Tier-0 supercomputers or hyperscaler cloud providers.

Artifacts that can be used to reproduce this paper's results are licensed under BSD 3-clause license and archived in \anonym{a}{an anonymous} Zenodo repository. \href{https://doi.org/10.5281/zenodo.18326325}{DOI: 10.5281/zenodo.18326326}

\section*{Acknowledgment}
We thank our colleagues Andreas Wolf, Tim Heldmann, Sebastian Kreutzer, and Tim Jammer for their helpful input and thoughtful comments during our discussions on the topic.

\renewcommand{\bibfont}{\footnotesize}
\printbibliography

@article{benoit_reducing_2018,
	title = {Reducing the energy consumption of large-scale computing systems through combined shutdown policies with multiple constraints},
	volume = {32},
	issn = {1094-3420},
	doi = {10.1177/1094342017714530},
	pages = {176--188},
	number = {1},
	journaltitle = {The International Journal of High Performance Computing Applications},
	author = {Benoit, Anne and Lefèvre, Laurent and Orgerie, Anne-Cécile and Raïs, Issam},
	year = {2018},
}

@online{SMARD,
  author = {Bundesnetzagentur},
  title = {Electricity market data},
  year = 2025,
  url = {https://www.smard.de/en/downloadcenter/download-market-data/},
  urldate = {2025-05-24}
}

@online{AEMO,
  author = {{Australian Energy Market Operator}},
  title = {Aggregated price and demand data},
  year = 2025,
  url = {https://www.aemo.com.au/energy-systems/electricity/national-electricity-market-nem/data-nem/aggregated-data},
  urldate = {2025-09-04}
}

@article{sartor_money_2010,
	title = {Money for Research, Not Energy Bills: Finding Energy and Cost Savings in High-Performance Computer Facility Designs},
	volume = {12},
	issn = {1558-366X},
	doi = {10.1109/MCSE.2010.137},
	shorttitle = {Money for Research, Not Energy Bills},
	pages = {11--19},
	number = {6},
	journaltitle = {Computing in Science \& Engineering},
	author = {Sartor, Dale and Wilson, Mark},
	date = {2010-11},
}

@article{silva_review_2024,
	title = {A review on the decarbonization of high-performance computing centers},
	volume = {189},
	issn = {1364-0321},
	doi = {10.1016/j.rser.2023.114019},
	pages = {114019},
	journaltitle = {Renewable and Sustainable Energy Reviews},
	author = {Silva, C. A. and Vilaça, R. and Pereira, A. and Bessa, R. J.},
	date = {2024-01-01},
}

@article{suarez_energy_2025,
	title = {Energy efficiency trends in {HPC}: what high-energy and astrophysicists need to know},
	volume = {13},
	issn = {2296-424X},
	doi = {10.3389/fphy.2025.1542474},
	journal = {Frontiers in Physics},
	author = {Suarez, Estela and Amaya, Jorge and Frank, Martin and Freyermuth, Oliver and Girone, Maria and Kostrzewa, Bartosz and Pfalzner, Susanne},
	date = {2025-04-28},
}

@article{martin_de_lagarde_how_2018,
	title = {How renewable production depresses electricity prices: Evidence from the German market},
	volume = {117},
	issn = {0301-4215},
	doi = {10.1016/j.enpol.2018.02.048},
	pages = {263--277},
	journaltitle = {Energy Policy},
	author = {Martin de Lagarde, Cyril and Lantz, Frédéric},
	date = {2018-06-01},
}

@article{paraschiv_impact_2014,
	title = {The impact of renewable energies on {EEX} day-ahead electricity prices},
	volume = {73},
	issn = {0301-4215},
	doi = {10.1016/j.enpol.2014.05.004},
	pages = {196--210},
	journaltitle = {Energy Policy},
	author = {Paraschiv, Florentina and Erni, David and Pietsch, Ralf},
	date = {2014-10-01},
}

@article{cincinelli_role_2025,
	title = {The role of geopolitical and climate risk in driving uncertainty in European electricity markets},
	volume = {144},
	issn = {0140-9883},
	doi = {10.1016/j.eneco.2025.108276},
	pages = {108276},
	journaltitle = {Energy Economics},
	author = {Cincinelli, Peter and Pellini, Elisabetta},
	date = {2025-04-01},
}

@article{tarraf_malleability_2024,
	title = {Malleability in Modern {HPC} Systems: Current Experiences, Challenges, and Future Opportunities},
	volume = {35},
	issn = {1558-2183},
	doi = {10.1109/TPDS.2024.3406764},
	pages = {1551--1564},
	number = {9},
	journaltitle = {{IEEE} Transactions on Parallel and Distributed Systems},
	author = {Tarraf, Ahmad and Schreiber, Martin and Cascajo, Alberto and Besnard, Jean-Baptiste and Vef, Marc-André and Huber, Dominik and Happ, Sonja and Brinkmann, André and Singh, David E. and Hoppe, Hans-Christian and Miranda, Alberto and Peña, Antonio J. and Machado, Rui and Garcia-Gasulla, Marta and Schulz, Martin and Carpenter, Paul and Pickartz, Simon and Rotaru, Tiberiu and Iserte, Sergio and Lopez, Victor and Ejarque, Jorge and Sirwani, Heena and Carretero, Jesus and Wolf, Felix},
	date = {2024-09},
}

@inproceedings{iserte_efficient_2017,
	location = {New York},
	title = {Efficient Scalable Computing through Flexible Applications and Adaptive Workloads},
	doi = {10.1109/ICPPW.2017.36},
	eventtitle = {46th International Conference on Parallel Processing Workshops ({ICPPW})},
	pages = {180--189},
	booktitle = {2017 46th {Internatioal} {Conference} {on} {Paralllel} {Processing} {Workshops} ({ICPPW})},
	publisher = {{IEEE}},
	author = {Iserte, Sergio and Mayo, Rafael and Quintana-Orti, Enrique S. and Beltran, Vicenc and Pena, Antonio J.},
	date = {2017},
    issn = {1530-2016},
}

@inproceedings{zhang_scheduling_2021,
	location = {Cham},
	title = {Scheduling Challenges for Variable Capacity Resources},
	doi = {10.1007/978-3-030-88224-2_10},
	pages = {190--209},
	booktitle = {Job Scheduling Strategies for Parallel Processing},
	publisher = {Springer International Publishing},
	author = {Zhang, Chaojie and Chien, Andrew A.},
	editor = {Klusáček, Dalibor and Cirne, Walfredo and Rodrigo, Gonzalo P.},
	date = {2021},
}

@article{chase_managing_2001,
	title = {Managing energy and server resources in hosting centers},
	volume = {35},
	issn = {0163-5980},
	doi = {10.1145/502059.502045},
	pages = {103--116},
	number = {5},
	journaltitle = {{SIGOPS} Oper. Syst. Rev.},
	author = {Chase, Jeffrey S. and Anderson, Darrell C. and Thakar, Prachi N. and Vahdat, Amin M. and Doyle, Ronald P.},
	date = {2001-10-21},
}

@inproceedings{pinheiro_load_2001,
	location = {Barcelona, Spain},
	title = {Load Balancing and Unbalancing for Power and Performance in Cluster-Based Systems},
	eventtitle = {2nd Workshop on Compilers and Operating Systems for Low Power},
    booktitle = {2nd Workshop on Compilers and Operating Systems for Low Power},
	author = {Pinheiro, Eduardo and Bianchini, Ricardo and Carrera, Enrique and Heath, Taliver},
	date = {2001-09-11},
}

@inproceedings{lin_adapting_2023,
	location = {New York, {NY}, {USA}},
	title = {Adapting Datacenter Capacity for Greener Datacenters and Grid},
	isbn = {979-8-4007-0032-3},
	doi = {10.1145/3575813.3595197},
	series = {e-Energy '23},
	pages = {200--213},
	booktitle = {Proceedings of the 14th {ACM} International Conference on Future Energy Systems},
	publisher = {Association for Computing Machinery},
	author = {Lin, Liuzixuan and Chien, Andrew A},
	date = {2023-06-16},
}

@article{casagrande_dont_2022,
	title = {Don’t hurry be green: scheduling servers shutdown in grid computing with deep reinforcement learning},
	volume = {13},
	doi = {10.1504/IJGUC.2022.128303},
	pages = {589--606},
	number = {6},
	journaltitle = {{International} {Journal} {of} {Grid} {and} {Utility} {Computing}},
	author = {Casagrande, Lucas Camelo and Koslovski, Guilherme Piêgas and Miers, Charles Christian and Pillon, Maurício Aronne and Gonzalez, Nelson Mimura},
	date = {2022},
	langid = {english},
}

@article{rais_quantifying_2018,
	title = {Quantifying the impact of shutdown techniques for energy-efficient data centers},
	volume = {30},
	rights = {Copyright © 2018 John Wiley \& Sons, Ltd.},
	issn = {1532-0634},
	doi = {10.1002/cpe.4471},
	pages = {e4471},
	number = {17},
	journaltitle = {Concurrency and Computation: Practice and Experience},
	author = {Raïs, Issam and Orgerie, Anne-Cécile and Quinson, Martin and Lefèvre, Laurent},
	date = {2018},
}

@article{lago_forecasting_2021,
	title = {Forecasting day-ahead electricity prices: A review of state-of-the-art algorithms, best practices and an open-access benchmark},
	volume = {293},
	issn = {0306-2619},
	doi = {10.1016/j.apenergy.2021.116983},
	shorttitle = {Forecasting day-ahead electricity prices},
	journaltitle = {Applied Energy},
	author = {Lago, Jesus and Marcjasz, Grzegorz and De Schutter, Bart and Weron, Rafał},
	date = {2021-07-01},
}

@inproceedings{bodner2025case,
  title={A Case for Ecological Efficiency in Database Server Lifecycles},
  author={Bodner, Thomas and Boissier, Martin and Rabl, Tilmann and Salazar-D{\'\i}az, Ricardo and Schmeller, Florian and Strassenburg, Nils and Tolovski, Ilin and Weisgut, Marcel and Yue, Wang},
  booktitle={Proc. 15th Annual Conference on Innovative Data Systems Research (CIDR 2025)},
  year={2025}
}

@article{CORNELL20241421,
title = {A probabilistic forecast methodology for volatile electricity prices in the Australian National Electricity Market},
journal = {International Journal of Forecasting},
volume = {40},
number = {4},
pages = {1421-1437},
year = {2024},
issn = {0169-2070},
doi = {https://doi.org/10.1016/j.ijforecast.2023.12.003},
author = {Cameron Cornell and Nam Trong Dinh and S. Ali Pourmousavi},
}

@article{chien_zero-carbon_2015,
	title = {The Zero-Carbon Cloud: High-Value, Dispatchable Demand for Renewable Power Generators},
	volume = {28},
	issn = {1040-6190},
	doi = {10.1016/j.tej.2015.09.010},
	shorttitle = {The Zero-Carbon Cloud},
	pages = {110--118},
	number = {8},
	journaltitle = {The Electricity Journal},
	shortjournal = {The Electricity Journal},
	author = {Chien, Andrew A. and Wolski, Richard and Yang, Fan},
	date = {2015-10-01},
}

@misc{electricitymaps,
  author       = {{Electricity Maps}},
  title        = {2024 Day-Ahead Electricity Price Data for Multiple Countries},
  date         = {2025-12-02},
  howpublished = {\url{https://www.electricitymaps.com}},
  note         = {Historical day-ahead electricity price data for multiple countries retrieved via the Electricity Maps API}
}

\end{document}